# Magnetotransport properties of ferromagnetic $Pr_{0.85}Ca_{0.15}MnO_3$/ferroelectric $Ba_{0.6}Sr_{0.4}TiO_3$ superlattice films

P. Murugavel, P. Padhan, and W. Prellier

Laboratoire CRISMAT, CNRS UMR 6508, ENSICAEN, 6Bd du Maréchal Juin,

F-14050 Caen Cedex, FRANCE.

## Abstract

Artificial superlattices designed with ferromagnetic $Pr_{0.85}Ca_{0.15}MnO_3$ insulating layer and ferroelectric $Ba_{0.6}Sr_{0.4}TiO_3$ layer were grown on (100) $SrTiO_3$ substrates. The magnetotransport properties were measured with the current perpendicular to the plane geometry. An increase in magnetoresistance (MR), with no significant low field effect, was observed as the number of ferroelectric $Ba_{0.6}Sr_{0.4}TiO_3$ layer thickness increases even up to 9 unit cells. For example, the superlattice $[(Pr_{0.85}Ca_{0.15}MnO_3)_{10}(Ba_{0.6}Sr_{0.4}TiO_3)_9]_{25}$ shows 35 % MR at 100 K, though the $Pr_{0.85}Ca_{0.15}MnO_3$ film was a robust insulator with negligible MR even at high applied magnetic field. This observed large MR cannot be explained by simple interfacial ferromagnetism or by the tunneling magnetoresistance. One possible explanation could be the effect due to the ferroelectric spacer layer and the associated magnetoelectric coupling.

In recent years, extensive investigations of perovskite superlattices have been reported,[1-5] because of its diverse and unusual transport as well as magnetic properties, which are not commonly observed in samples prepared by classical solid state chemistry route. In addition, there is a great interest on colossal magnetoresistance (MR) properties observed in manganite-based perovskite materials. Thus, several interesting superlattice structures made by the combinations among ferromagnetic, antiferromagnetic, paramagnetic insulator, and paramagnetic metal, yielded unusual physical properties[1-7] along with enhanced magnetoresistance (MR). The enhanced MR at low temperatures in these superlattices were attributed mostly to the existence of various interesting physical phenomenon occurring at the interfaces, for example, canted spin arrangements of ferromagnetic layer in $La_{0.6}Sr_{0.4}MnO_3/La_{0.6}Sr_{0.4}FeO_3$[2] and $La_{0.6}Sr_{0.4}MnO_3/SrTiO_3$[4] superlattice, formation of interfacial ferromagnetism in $CaMnO_3$/ $CaRuO_3$[5] superlattice, and spin disordered region in $La_{0.7}Ca_{0.3}MnO_3/SrTiO_3$[6] superlattice. The significant value of MR were all observed at very low temperatures and its changes were strongly associated with the change in its magnetic properties. The manganites-based multilayers also seemed to show low field tunneling MR at very low temperature[7]. However, an another interesting superlattice structure composed of alternate stacking of ferromagnetic and ferroelectric materials, and its physical properties have rarely been investigated.

Moreover, recent study of mutiferroic materials,[8,9] *i.e.* materials showing simultaneous magnetic, ferroelectric, and/or ferroelastic properties, in thin film form renewed further interest in the ferromagnetic/ferroelectric superlattices because of its technological importance. In this letter, we present our results on the electrical and magnetic transport properties of one such superlattice structures built from ferromagnetic $Pr_{0.85}Ca_{0.15}MnO_3$ (PCMO) layer, whose ferromagnetic transition temperature ($T_C$) is nearly at 130 K, and ferroelectric $Ba_{0.6}Sr_{0.4}TiO_3$ (BST) layer, whose ferroelectric transition temperature is nearly at 260 K. The transport properties of our series of superlattices consisting of $(PCMO_{10}/BST_N)_{25}$ (where 10 and N indicates the number of PCMO and BST unit cells, respectively) show an increase in magnetoresistance (MR) in high magnetic field with further enhancement in its value, by increasing the ferroelectric spacer layer thickness. We believe that the observed enhancement in magnetoresistance could be related to the effect due to ferroelectric spacer layer and the magnetoelectric coupling associated to the multiferroic materials.

Superlattices of $(PCMO_{10}/BST_N)_{25}$ ($N$=1-9) were fabricated on (100) $SrTiO_3$ (STO) substrates by a multitarget pulsed-laser deposition method using KrF laser (laser wavelength $\lambda$=248 nm, 200 mJ) focussed on PCMO and BST polycrystalline targets. The targets were prepared by conventional solid state route using powders of $Pr_2O_6$, $CaCO_3$, $MnO_2$, $BaCO_3$, $SrCO_3$, and $TiO_2$. All the films were deposited at 720 °C at 100 mTorr of oxygen pressure. The calibrated deposition rate was 0.272 and 0.367 Å/pulse for PCMO and BST, respectively. After the deposition, the films were cooled under 300 Torr of oxygen pressure at the rate of 13 °C/min down to room temperature. In all our films, the thickness of each PCMO layer was fixed at 10 unit cell (u.c.), and the BST layer was varied from 1 to 9 u.c.. The superlattice is composed of 25 repeated units of $(PCMO_{10}/BST_N)$ bilayers with PCMO as the bottom layer and BST as the top layer. After the deposition, the samples were characterized by x-ray diffraction (XRD) using Seifert 3000P diffractometer (Cu K$\alpha$, $\lambda$=1.5406Å). The resistance ($R$) was measured on selected samples prepared in a special geometry, using a four-probe method, by applying the current perpendicular to the plane (CPP). A magnetic field (0-7 T), parallel to the film surface, was applied using a superconducting magnet. Magnetization ($M$) was measured as a function of temperature ($T$) and field ($H$) using a superconducting quantum interference device magnetometer (SQUID)

In Fig. 1a, we show the $\theta$-$2\theta$ XRD scan around the (002) fundamental peak (42° - 50° in $2\theta$) of various superlattices. The number $N$ of the BST layer in the [10/$N$] superlattice was determined by XRD measurement of superlattice peaks. As the BST layer thickness decreases from 9 u.c. the fundamental diffraction peak of the superlattice (indicated as 0 in the Fig. 1a), shifted towards the (002) STO peak and overlaps it at lower $N$ value. The denoted number $i$ indicates the *ith* satellite peak. The presence of higher order satellite peaks adjacent to the main peak, arising from chemical modulation of multilayer structure, indicates that the films were indeed coherent heterostructurally grown. We have carried out the XRD simulation of the superlattice structure using *DIFFaX* program[10] and it is found that the experimentally measured peaks are in good agreement with the simulated one. The full-width-at-half-maxima (*FWHM*) of the rocking curve, recorded around the fundamental (002) diffraction peak of the superlattice samples of various $N$ (Fig.1b), are very close to the instrumental broadening (<0.3), indicating a high crystalline quality and a good coherency. We also show the superlattice period $\Lambda$ for different $N$ in Fig.1b. [The value of $\Lambda$ were calculated from, $\Lambda = \lambda/(2\times(\sin\theta_i - \sin\theta_{i+1}))$, where $\lambda$ is the x-ray wavelength, $\theta_i$, and $\theta_{i+1}$ are the angular position of the *ith* and *(i+1)th* order satellite peak[11]]. The linear variation of $\Lambda$ with $N$ is an additional evidence for the good quality of the films.

We have measured the MR (MR=100 $\times$ ($R_H$ – $R_0$)/$R_0$, where $R_H$ and $R_0$ is the resistance measured with and without magnetic field, respectively) in a CPP geometry at 100 K and the results are shown in Fig. 2a. The Fig. 2b shows the *R(T)* curve of the superlattices of 10/2, 10/5, and 10/9. For the CPP measurements the samples were prepared in a special geometry with the conductive $LaNiO_3$ electrode in L-like shape (see the inset in Fig. 2a). Using L shaped electrode, we could control the junction area by slightly displacing the top electrode relative to the bottom one in diagonal direction. The junction prepared in this way allows the junction current to follow a straight line from one electrode to another, thus avoiding geometrical effect commonly arising in junctions prepared in usual cross-strip geometry. The results are shown for the superlattices of 10/2, 10/5, and 10/9. All these films are showing an increase in MR with increase in field up to the maximum applied field of 7 T. We did not see any significant low field MR in these films. Interestingly, the MR value is further enhanced with increase in ferroelectric spacer layer thickness. For example, the samples 10/2, 10/5, and 10/9 show 12.5, 21.8, and 35.3% MR at 7 T, respectively. Note that the $Pr_{1-x}Ca_xMnO_3$ ($0.15<x<0.25$) is one of the robust ferromagnetic insulator both in bulk and thin films form with a very small change in resistance with applied magnetic field[12]. Thus, the observed high value of MR (35 % at 7 T) even at 100 K with its enhancing behavior with higher ferroelectric layer thickness is an important result to be noted.

Tunneling magnetoresistance (TMR) reported in superlattices of ferromagnetic/paramagnetic insulator,[7] where the MR is observed usually up to 2 to 3 unit cell spacer layer thickness, could provide a possible explanation for our MR results. However, we observed high MR value even in spacer layer thickness of $N$=9 and its value is nearly an order of magnitude higher than the PCMO film (note that the layers in our superlattices are highly insulators). For comparison the MR of the pure PCMO film with same thickness as in superlattice films (25 × 10 u.c) is also shown in Fig. 2a ($N$=0 curve). The MR of pure PCMO is negligibly small (4.2% at 7 T) compared to the superlattices. Thus, the observed MR in our superlattices cannot be explained by TMR. Another possible explanation could be the reported interfacial ferromagnetism, which is usually associated with significant change in saturation magnetization $M_S$ and $T_C$.[3-6] To verify it, we have measured the magnetization with field and temperature, and the results are shown in Fig. 3. As a representative example, Fig. 3a shows the magnetic hysteresis loop of $(PCMO_{10}/BST_5)_{25}$ sample measured at 100 K and the inset show the magnetization with temperature plot. Fig. 3b shows the change in $M_S$ and $T_C$ with spacer layer thickness. From Fig. 3a, it is clearly seen that the magnetic moment saturate at a magnetic field below 0.5 T and shows no further increase with $H$. However, the MR of the superlattice exhibits an increasing trend even at high applied magnetic field indicating no clear relations between the magnetization and MR. Also Fig. 3b indicate that the changes in $M_S$ and $T_C$ of our films are not significantly high with increase in spacer layer thickness. As a consequence, the interfacial ferromagnetism cannot be argued as a reason for our observed high MR.

Since our films are multilayers made of ferromagnetic and ferroelectric layers, the magnetoelectric coupling associated with the multilayers of magnetostrictive and piezoelectric perovskite oxides[9] could provide a possible explanation for our result. It has been reported that $Pr_{0.8}Ca_{0.2}MnO_3$ have negative differential resistance with electric field below its ferromagnetic transition temperature.[13] The electric field can be generated by ferroelectric BST layers in the superlattice structure by producing the charges at the interface. Thus, the increase in MR with BST layer thickness can be explained by the increase in charges at the interface which in turn changes the resistance of PCMO layer. Similarly, the increase in MR with magnetic field can be attributed to the negative differential resistance due to the charges developed at the interfaces by the piezoelectric effect of BST layer induced by the magnetostriction associated with the PCMO layer[9]. However, further experiments are needed to clarify this effect.

In conclusion, we have made good quality superlattice of ferromagnetic/ferroelectric materials, $Pr_{0.85}Ca_{0.15}MnO_3/Ba_{0.6}Sr_{0.4}TiO_3$. The transport properties of the samples measured in CPP direction at 100 K shows an increase in magnetoresistance up to the maximum applied magnetic field of 7 T. No significant low field magnetoresistance were found. The magnetoresistance of the superlattice increases with increase of the ferroelectric spacer layer thickness. We calculated a MR of 35.3% at 100 K for the N=9 sample which is nearly an order of magnitude higher than the $Pr_{0.85}Ca_{0.15}MnO_3$ film. The high value of MR and its increasing trend with increase ferroelectric layer thickness, is an important result to note which cannot be attributed either to the interfacial ferromagnetism or to the tunneling magnetoresistance. The possible explanation for the MR could be the change in resistance associated with the electric charges developed at the interface due to the ferroelectric $Ba_{0.6}Sr_{0.4}TiO_3$ layer and the magnetoelectric coupling effect in the superlattices. It would be interesting to study the dielectric properties of the superlattices under magnetic field to clarify the exact role of magnetoelectric coupling on the magnetotransport properties of these materials, and the investigations are on in this direction.

Acknowledgements:

One of the author (P.M.) acknowledges the Ministere de la Jeunesse et de l'Education Nationale for his fellowship (2003/87). We also greatly acknowledge the financial support of Centre Franco-Indien pour la Promotion de la Recherche Avancee/Indo-French Centre for the Promotion of Advance Research (CEFIPRA/IFCPAR) under Project N° 2808-1.

Figure Captions:

FIG. 1. (a) Observed (dark line) and calculated (less dark line) $\theta$–$2\theta$ XRD scan recorded around the (002) reflection of $SrTiO_3$ substrate for various superlattices $(PCMO_{10}/BST_N)_{25}$ ($N$=1-9). The symbol $i$ indicate the number of satellite peak. (b) Evolution of the *FWHM* and the superlattice period ***L*** as a function of the spacer layer thickness (N).

FIG. 2. (a) MR at high magnetic field measured at 100 K for the superlattices $[(PCMO)_{10}/(BST)_N]_{25}$ of $N$=0, 2, 5, and 10. The inset shows the junction structure used for the measurement. (b) Temperature dependent resistance for the superlattices $(PCMO)_{10}/(BST)_N]_{25}$ of $N$=0, 2, and 10. In all the measurement the current is applied perpendicular to the plane of the sample.

FIG.3. (a) Magnetic hysteresis loop measured at 100 K for $(PCMO_{10}/BST_5)_{25}$ superlattice. The inset shows the corresponding temperature-dependent magnetization curve. (b) Evolution of the saturated magnetization ($M_S$) and the Curie temperature ($T_C$) as a function of the spacer layer thickness ($N$)
.

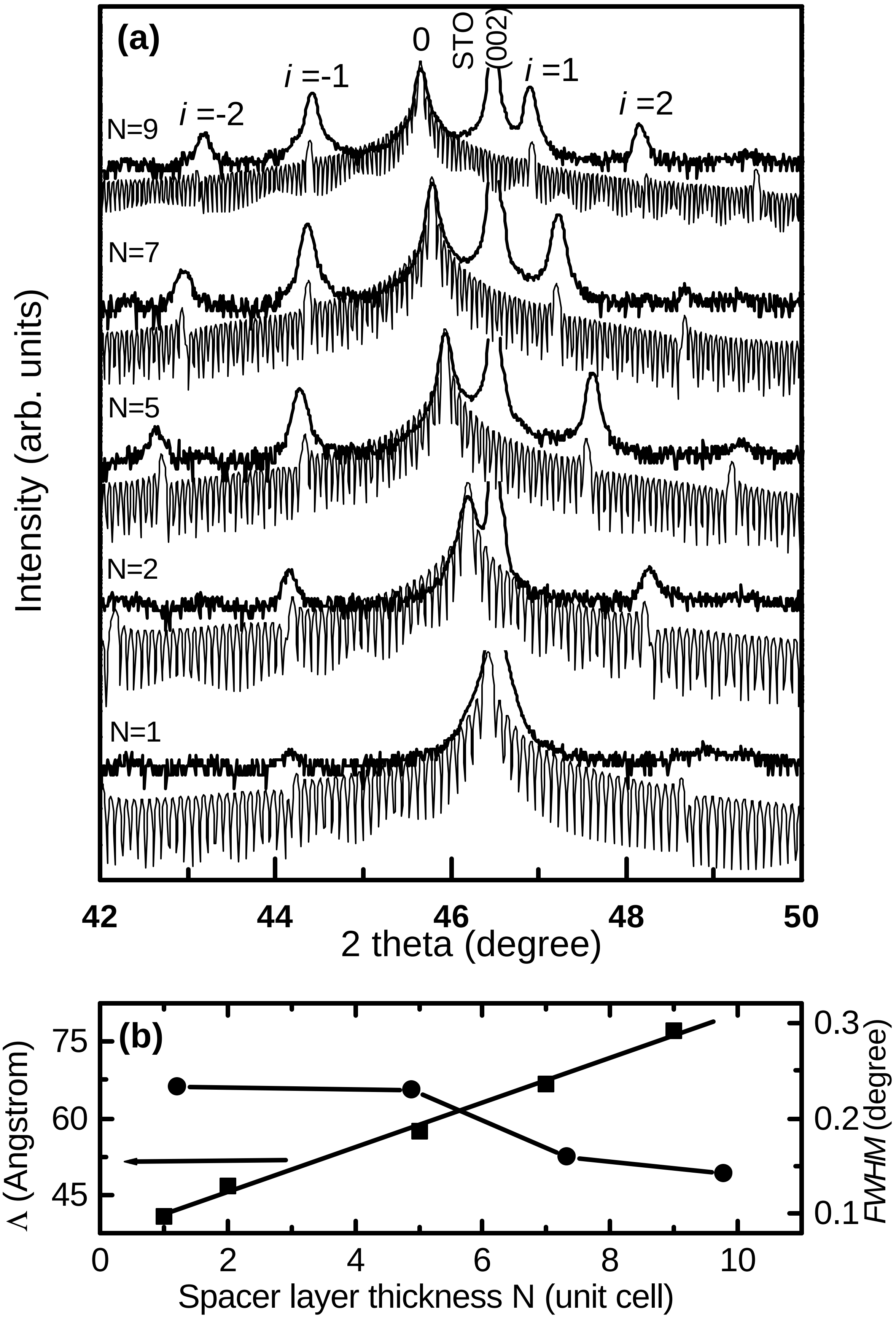

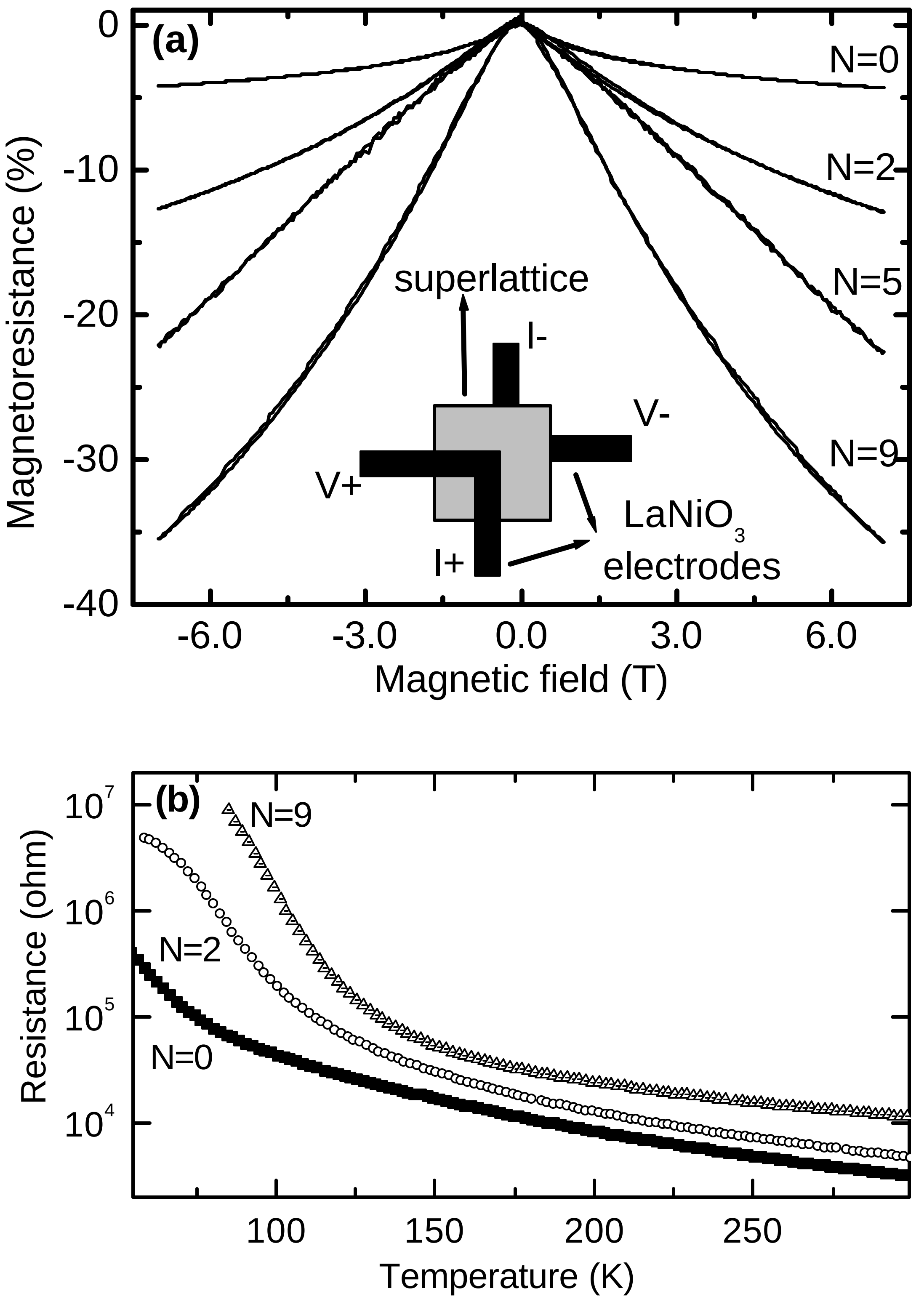

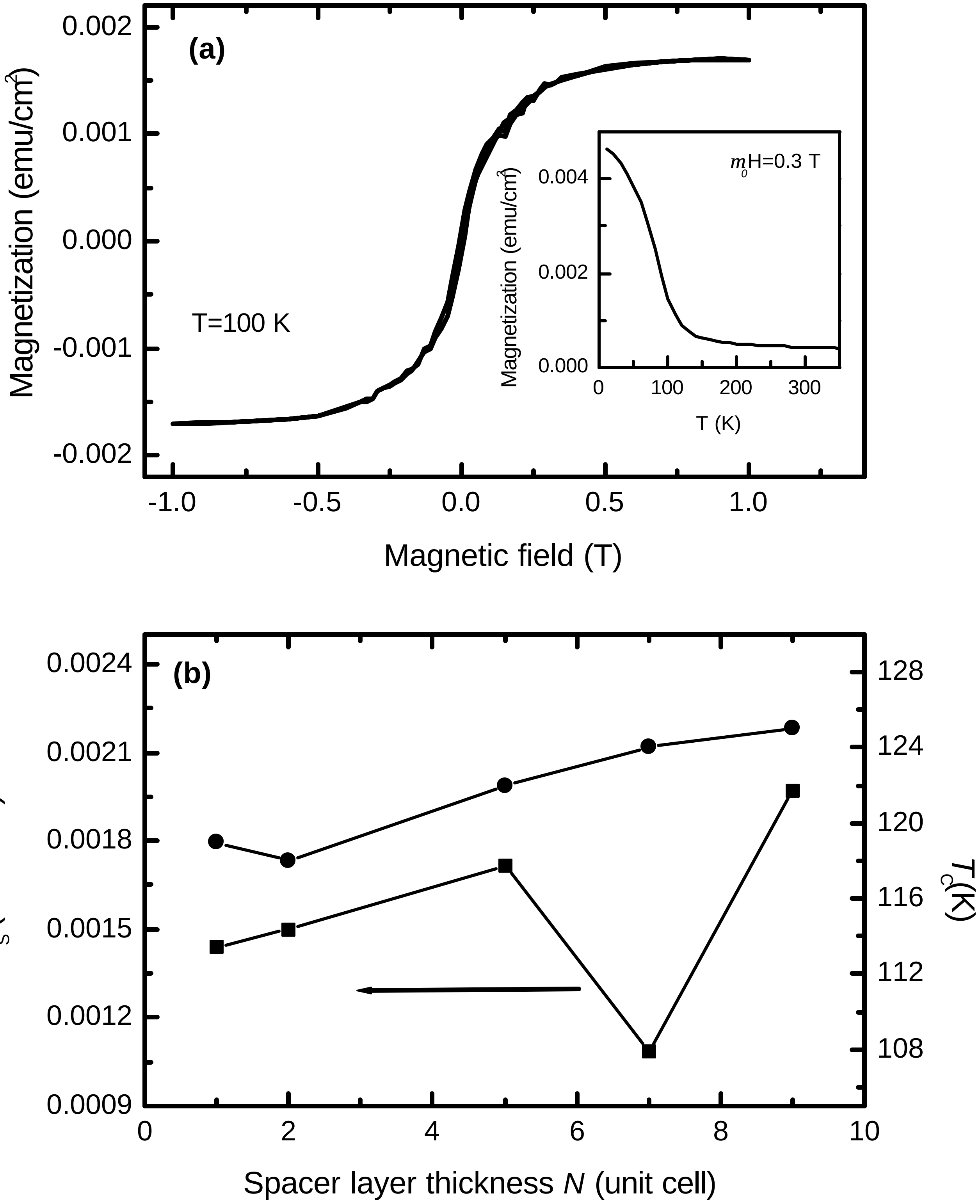